\documentclass[aps,prl,reprint,showpacs,floatfix,twocolumn]{revtex4}
\usepackage{graphicx}

\begin{document}

\title{A quantum detector for photon entanglement}

\author{J. Huwer$^{1,2}$\footnote{Corresponding author. \\ Electronic address:
jan.huwer@physik.uni-saarland.de}, J. Ghosh$^{1,2}$, N. Piro$^{1}$\footnote{Current address: Ecole Polytechnique F\'ed\'erale de Lausanne (EPFL), 1015 Lausanne, Switzerland.}, M. Schug$^{2}$, F.
Dubin$^{1}$, J. Eschner$^{1,2}$}

\affiliation{
$^1$ICFO -- The Institute of Photonic Sciences, Av. Carl Friedrich Gauss 3, 08860 Castelldefels (Barcelona), Spain\\
$^2$Universit\"at des Saarlandes, Experimentalphysik, Campus E2 6, 66123 Saarbr\"ucken,
Germany}

\date{\today}

\begin{abstract}
We use a single trapped $^{40}$Ca$^+$ ion as a resonant, polarization-sensitive absorber
to detect and characterize the entanglement of tunable narrowband photon pairs from a
spontaneous parametric down-conversion source. Single-photon absorption is marked by a
quantum jump in the ion and heralded by coincident detection of the partner photon. For
three polarization basis settings of absorption and detection of the herald, we find
maximum coincidences always for orthogonal polarizations. The polarization entanglement
is further evidenced by tomographic reconstruction of the biphoton quantum state.
\end{abstract}


\maketitle

The realization of quantum networks is a primary goal in quantum information science
\cite{Zoller2005}, remote entanglement being the basic resource required to establish
quantum channels between their nodes \cite{Kimble2008}. Promising schemes for
experimental implementations are based on single atomic qubits as nodes, where quantum
information is stored and processed, and single photons to communicate between the nodes
and generate entanglement. Therefore the very basic requirement is to be able to
efficiently and reversibly transfer quantum states between atoms and photons at the
single particle level. In that sense, controlled emission and absorption of single
photons by a single atomic particle are key enabling tools for quantum optical
information technologies \cite{Cirac1997, Monroe2002}.

In studies of single-photon emission, achievements include control of the shape,
frequency, polarization, and bandwidth of the generated single photons \cite{Keller2004,
Legero2004, Wilk2007PRL, McKeever2004, Maunz2007, Barros2009, Almendros2009}, and the
demonstration of atom-photon entanglement \cite{Blinov2004, Volz2006, Wilk2007,
Blatt2012}. Based on the latter, distant entanglement of single atoms has been
established \cite{Moehring2007} and employed \cite{Maunz2009} by the projective
measurement of two photons, each entangled with its emitting atom. A fully bi-directional
atom-photon interface implies the transfer of quantum states also by controlled
\emph{absorption} of a single photon \cite{Cirac1997}. Then the entanglement shared by a
photon pair as they come from a spontaneous parametric down-conversion (SPDC) source
could be transferred to two distant nodes of the network, entangling them with each other
\cite{Lloyd2001}. Experimental progress in this direction includes designing optical
systems that optimize the atom-photon coupling in free space \cite{Sondermann2007,
Wrigge2008, Maiwald2009, Streed2011, VanDevender2010}, detecting the attenuation and
phase shift of a weak laser beam by a single atom \cite{Tey2008, Aljunid2009}, and
transferring photons between atoms in cavities \cite{Rempe2011, Rempe2012}.

Controlling both emission and absorption will pave the way towards implementing quantum
networking scenarios, where transmission of quantum information across the network is
directly linked with its local processing in atoms \cite{Monroe2002, Luo2009, Duan2010}.
In this context, single trapped ions provide optimal conditions for quantum information
processing, meeting the requirements \cite{DiVincenzo2000} of high-fidelity state
manipulation and detection schemes, as well as the controlled interaction of several
qubits \cite{SchmidtKaler2003, Leibfried2003, Blatt2008}. At the same time, entangled
SPDC photon pairs offer robust and simple generation of high-purity entanglement at large
rate and thereby serve as an optimal resource for communication over quantum channels
\cite{Bouwmeester1997, Weihs1998, Jennewein2000, Ursin2007}. In our previous work
\cite{Schuck2010, Piro2011} we operated for the first time a hybrid quantum system
integrating single trapped ions and SPDC photon pairs. We demonstrated absorption of
single photons by the ion \cite{Schuck2010} and showed that the absorption of one photon
from a pair can be heralded by the coincident detection of its partner photon
\cite{Piro2011}; time and frequency correlation of the photon pair are manifested in the
heralded absorption process.

Here we report an experiment where the polarization entanglement of the SPDC photon pairs
is manifested through quantum tomography measurements, using the single ion as a
polarization-sensitive single-photon detector, prepared by optical pumping in a suitably
oriented magnetic field.
We find a fidelity of 93\% to the maximally entangled Bell state of the photon pairs.
This opens up the perspective of heralded storage of photon polarization in a single ion.

A schematic representation of our experimental setup is displayed in Fig.~1. A single
$^{40}$Ca$^+$ ion is confined and laser cooled in a linear Paul trap which is placed
between two high numerical aperture laser objectives (HALOs) \cite{Gerber2009}. The ion
is addressed by various laser beams for cooling and optical pumping. The HALOs serve for
efficiently collecting the ion's laser-excited fluorescence on a photomultiplier (PMT)
and for coupling the ion to single photons in a single optical mode. The SPDC photon
source consists of a periodically poled, type II phase-matched KTP crystal which is
continuously pumped at 427~nm. It emits polarization-entangled photon pairs in a 200~GHz
wide spectral band centered at the frequency of the D$_{5/2}-$P$_{3/2}$ electronic
transition of $^{40}$Ca$^+$ at 854~nm \cite{Haase2009, Piro2009}. To select entangled
pairs which are resonant with the ion, the pairs are split by a non-polarizing beam
splitter (BS) and then one output mode is filtered by two cascaded Fabry-Perot cavities
matching the atomic transition in center frequency and bandwidth (22~MHz). Before the
filtering stage, the photons pass through a polarization analyzing unit, comprising of a
quarter wave plate (QWP), a half wave plate (HWP) and a polarizing beam splitter (PBS).
Photons that pass through the filter ("trigger photons") are detected by an avalanche
photo diode (APD). The unfiltered photons in the other output mode of the BS are coupled
to the ion through one of the HALOs. Fiber polarization controllers are used to
neutralize polarization rotations that occur inside the 16~m long single-mode fiber
connecting the two setups. Resonant absorption of a photon by the ion is coincident with
(and hence heralded by) trigger photon detection with about 7\% probability
\cite{Piro2011}.

\begin{figure}[t]
\centering
\includegraphics[width=0.48\textwidth]{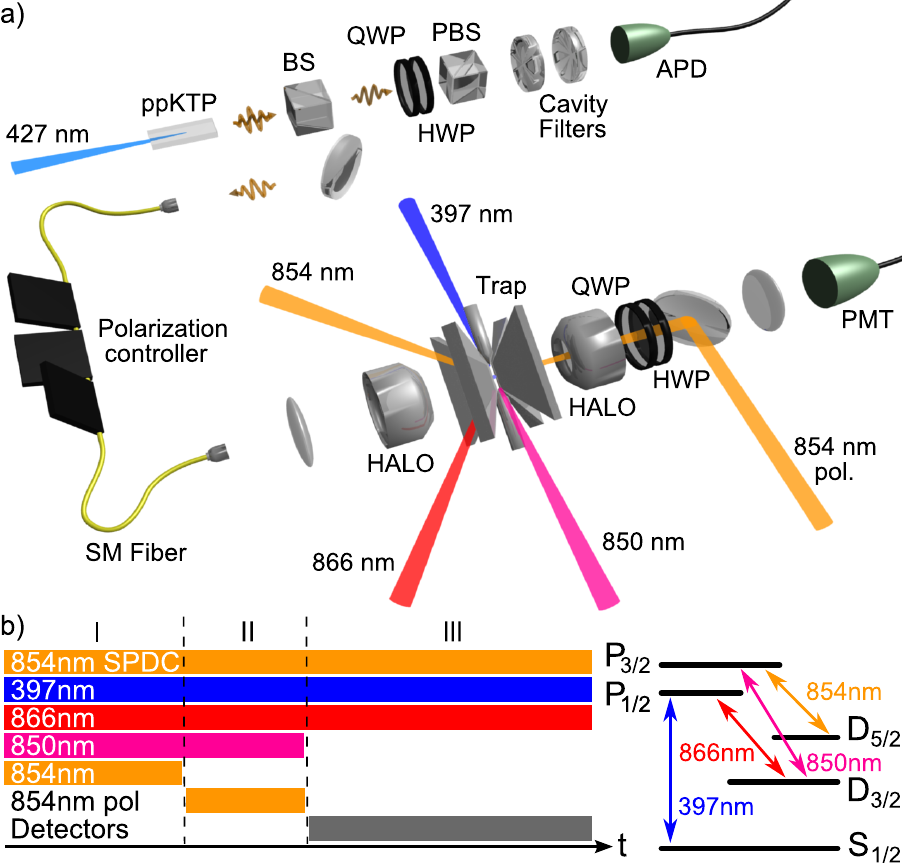}
\caption{(Color online) (a) Experimental setup consisting of SPDC photon source (top) and
ion trap (bottom). (b) Laser excitation scheme for ion-photon interaction and
corresponding levels and transitions in $^{40}$Ca$^+$. For details see text.}
\end{figure}

The photon source produces maximally entangled photon pairs in the Bell singlet state
$\left|\Psi^{-}\right\rangle$, in which the polarizations of the two photons are
anti-correlated in all polarization bases. In the measurement we use the \emph{R-L}
(right-left circular), \emph{H-V} (horizontal-vertical linear), and \emph{D-A}
(diagonal-antidiagonal linear) basis pairs. Transfer of the photon polarization to the
ion in a certain basis is obtained if the ion absorbs one basis state but not the other.
Photon entanglement is therefore manifested
by maximum coincidence of polarization-sensitive absorption and trigger photon detection in orthogonal polarization
states.

Fig.~1b presents the laser excitation sequence used to control the ion's interaction with
a single down-conversion photon. (I) Each period starts with Doppler cooling of the ion motion to
the Lamb-Dicke limit (wave packet expansion $\ll$ laser wavelength) with laser light at
397~nm and 866~nm. (II) Thereafter, the internal state of the ion is prepared for
polarization-sensitive absorption of SPDC photons: laser light at 850~nm populates the
D$_{5/2}$ level while at the same time, laser light at 854~nm optically pumps the ion
into specific Zeeman sub-levels of that manifold. The prepared state is controlled by the
polarization ${\bf E}$ and propagation direction ${\bf k}$ of the 854~nm pumping light
with respect to the orientation of the magnetic field ${\bf B}$ that sets the
quantization axis: with $\sigma$-polarized light on axis with the magnetic field (${\bf
k} \parallel {\bf B}$) we prepare a mixture of the states either with magnetic moments
$m=\left\{+\frac{3}{2},+\frac{5}{2}\right\}$, or with
$m=\left\{-\frac{3}{2},-\frac{5}{2}\right\}$. With $\pi$-polarized light propagating
perpendicular to the magnetic field orientation (${\bf k} \perp {\bf B} \parallel {\bf
E}$), we prepare an incoherent mixture of the states with
$m=\left\{-\frac{5}{2},+\frac{5}{2}\right\}$. In all cases, after preparation, an SPDC photon with the same
polarization as the preparing light cannot be absorbed, while the orthogonal polarization leads
to maximum absorption. (III) Finally, in the detection phase of the sequence, the ion is
exposed to the unfiltered SPDC photons, while the PMT and APD photodetectors are
activated. An absorption event during that phase is signaled by the onset of
fluorescence: transfer of the electronic population from the D$_{5/2}$ to the P$_{3/2}$
manifold and subsequent spontaneous decay into the S$_{1/2}$ level (with 94$\%$ branching
ratio) leads to steady 397~nm fluorescence recorded by the PMT photodetector.
The sequence runs with a repetition rate of roughly 10~Hz.

The rate of coincidences for a given polarization setting is evaluated by computing the
second-order time correlation function $g^{(2)}(\tau)$ between the trigger photon
detection events on the APD and the very first fluorescence photon detected on the PMT,
when the ion started to emit fluorescence photons in phase III of the sequence \cite{Piro2011}. A peak at
time delay $\tau=0$ signals the correlation of absorption and trigger photon detection;
in absence of correlation the background rate is measured.

\begin{figure}[htbp]
\centering
\includegraphics[width=0.48\textwidth]{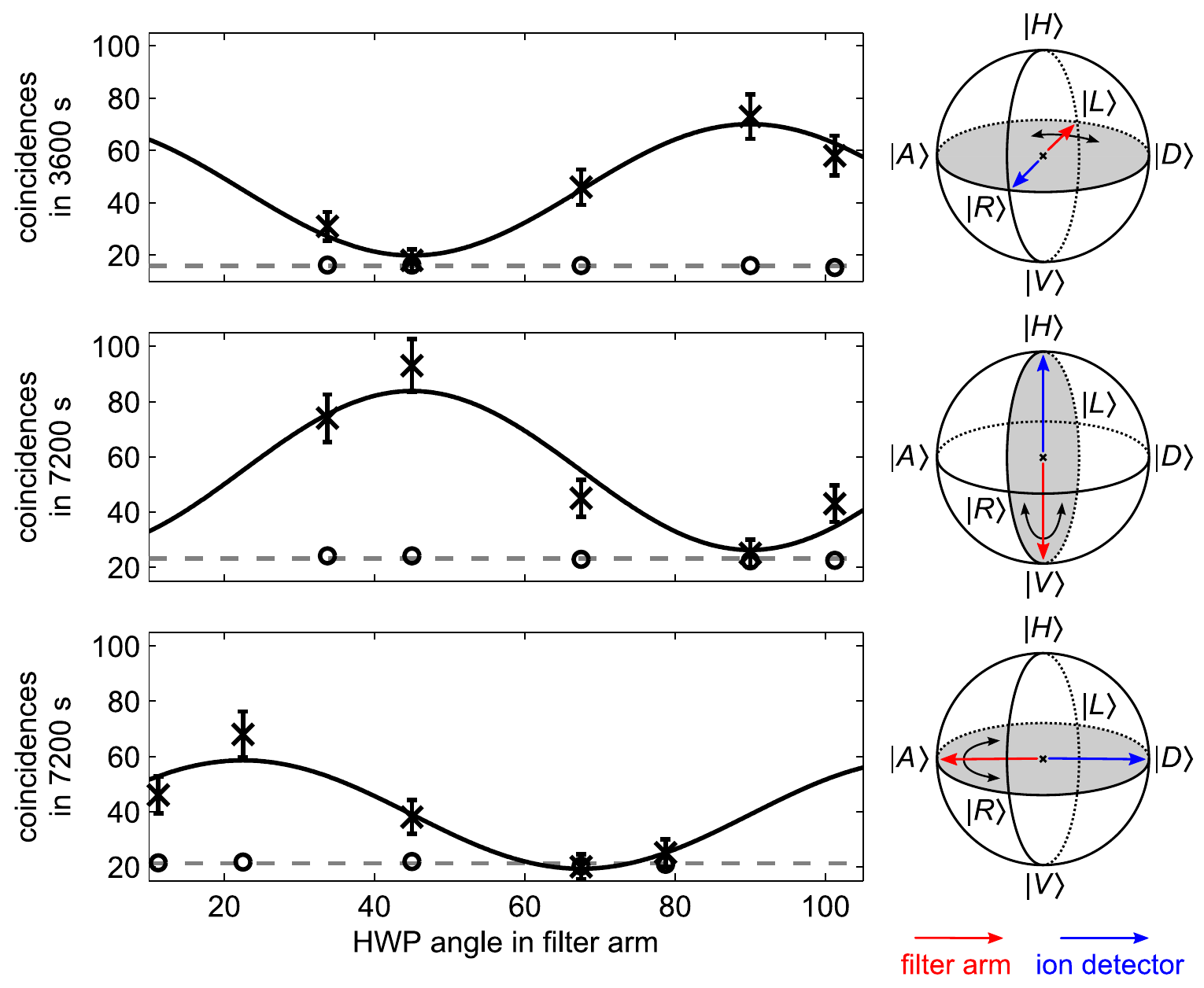}
\caption{ (Color online) Dependence of absorption-trigger coincidences on trigger photon
polarization for 3 different polarization bases, \emph{R-L} (top), \emph{H-V} (middle),
and \emph{D-A} (bottom). Data points (crosses) are extracted at $\tau=0$ from the
absorption-trigger correlation function $g^{(2)}(\tau)$ on a 10~$\mu$s time grid; the
corresponding background (circles), produced by accidental coincidences, is the average
over the whole $g^{(2)}$ function. Error bars correspond to one standard deviation
assuming Poissonian counting statistics. The curves are sinusoidal fits with a fixed period and offset angle. In the right-hand column we show the Poincar\'{e}
sphere with the setting of the ion (blue) and of the trigger photon detector (red, with
black arrows indicating variation by rotating the HWP).
 } \label{fig2}
\end{figure}

A measurement in the \emph{R-L} basis corresponds to preparing the ion in the
$m=\left\{+\frac{3}{2},+\frac{5}{2}\right\}$ or
$\left\{-\frac{3}{2},-\frac{5}{2}\right\}$ Zeeman levels, sending the SPDC photons with
${\bf k} \parallel {\bf B}$, and setting the waveplates in the filter arm such that
circularly polarized trigger photons are detected. Rotation of the HWP in that arm then
controls the relative orientation of the two circular polarizations. It should be
emphasized that the polarization of the photons sent to the ion is not changed, neither
filtered in this measurement. Fig.~2 (top) shows the observed number of coincidences for
different HWP angles. We observe a maximum of around 1 coincidence/min (73 coincidences
with 15 background events in 60~min) and a minimum within the background level for
orthogonal and same polarizations, respectively, which reflects the expected
anti-correlation of the photon polarizations. From a sinusoidal fit we find $56(6)\%$
visibility before background subtraction.

For measuring the correlations in the \emph{H-V} basis, optical pumping into the outer
Zeeman sub-levels $m=\left\{-\frac{5}{2},+\frac{5}{2}\right\}$ is applied, for which the
magnetic field is rotated by $90^\circ$, pointing now upwards in Fig.~1a. The SPDC
photons propagate with ${\bf k} \perp {\bf B}$ and can be absorbed if they are
\emph{H}-polarized but not if they are \emph{V}-polarized.
The trigger photons are detected with linear polarization along a direction varied by the HWP.
The result, as displayed in Fig.~2 (middle), exhibits a maximum of around 0.56
coincidences/min (92 with 24 background in 120~min) for \emph{H-V} and a minimum at the
background level for \emph{H-H}. The visibility derived from the sinusoidal fit is
$52(11)\%$.

To finally verify the presence of polarization correlations also in the \emph{D-A} basis,
preparation of the ion for absorption of linear polarization is used as before. To
achieve the $45^\circ$ rotation relative to the \emph{H-V} basis, in this case the
incoming photon state is rotated, using the fiber polarization controllers. The
waveplates in the filter arm are set accordingly and we measure coincidences, varying the
observed trigger polarization from \emph{D} to \emph{A} with the HWP. The result is
plotted in Fig.~2 (bottom). The maximum rate of coincidences is 0.4/min (67 with 21
background in 120~min) for \emph{D-A}, and for \emph{D-D} the rate reduces to the
background level. The sinusoidal fit gives a fringe visibility of $50(9)\%$ before
background subtraction. In all these measurements, the visibility is mainly determined by
the background events, arising due to heralds with lost partners and spontaneous decay
from the D$_{5/2}$ manifold. Nevertheless, the minimum points of the fringes are always,
within the experimental error, at the independently determined background level.

Having shown that the ion may be used as a polarization analyzer in the 3 principal bases
spanning the Poincar\'{e} sphere, we now apply the standard procedure of quantum
tomography \cite{James2001}. This allows reconstructing the density matrix of the
two-photon quantum state. To this end we perform coincidence measurements in 16
independent basis combinations, including now different basis settings at the ion and in
the filter arm. Using a maximum-likelihood state estimation, we derive the density matrix
$\rho$ shown in Fig.~3. In this analysis, we use the pure number of coincidences after
subtraction of the background counts.
From $\rho$, we derive the overlap fidelity with the maximally entangled singlet state $F
= \langle {\Psi}^{-} | \rho | {\Psi}^{-} \rangle = 0.93(4)$, as well as the concurrence
$C = 0.93(6)$ and tangle of formation $T = 0.86(11)$.

\begin{figure}[hbt]
\includegraphics[width=0.48\textwidth]{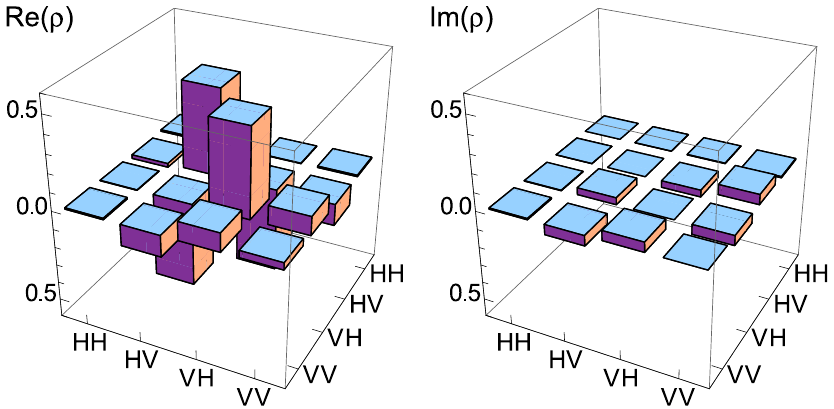}
\caption{(Color online) Real and imaginary parts, plotted in the \emph{H-V} polarization
basis, of the reconstructed density matrix $\rho$.}
\end{figure}

In summary, we achieve controlled interaction of a single atom with entangled
photons from a photon pair source, whereby the time correlation of the photon pairs is
employed for heralding the absorption process. Preparing the single atom as a
polarization-sensitive absorber in the three principal bases, we demonstrate that the
polarization entanglement of the photon pairs is manifested in the absorption-herald
correlation; this is confirmed by performing a full tomography of the quantum state. More
generally, we use a single isolated quantum system, a trapped ion, to characterize the
properties of another fundamental quantum system, entangled photons. In the context of
quantum information technology this is a necessary step for implementing a bi-directional
atom-photon quantum interface at the single particle level.
Transfer of the photonic entanglement to single atoms additionally requires coherent
preparation before absorption and measurement of the photon emitted after the absorption
event, for which schemes have been proposed \cite{Rohde2011}.

\begin{acknowledgments}
We acknowledge support by the European Commission (Integrating Projects SCALA, AQUTE; EMALI,
MRTN-CT-2006-035369) and the Spanish MICINN (QOIT, CSD2006-00019; QLIQS, FIS2005-08257;
QNLP, FIS2007-66944; CMMC, FIS2007-29999-E). J.G. is an Alexander von Humboldt fellow.
\end{acknowledgments}

{}

\end{document}